\documentclass[showpacs,twocolumn,floatfix,amsmath,amssymb]{revtex4}
\usepackage{graphicx,amsmath,color,amsmath,amssymb}

\usepackage{graphicx,amssymb}
\usepackage[T1]{fontenc}
\usepackage[latin9]{inputenc}
\usepackage{amsmath}

\usepackage{color}

\newcommand{\1}{\begin{equation}}
\newcommand{\2}{\end{equation}}
\newcommand{\ea}{\begin{eqnarray}}
\newcommand{\ee}{\end{eqnarray}}
\newcommand{\bea}{\begin{eqnarray}}
\newcommand{\eea}{\end{eqnarray}}
\newcommand{\bee}{\begin{eqnarray*}}
\newcommand{\eee}{\end{eqnarray*}}

\newcommand{\comm}[2]{\left[#1\,,\,#2\right]}

\newcommand{\der}[1]{{\frac{\rm d}{{\rm d}#1}}}
\newcommand{\dnd}[2]{{\frac{{\rm d}#2}{{\rm d}#1}}}
\newcommand{\pd}[2]{{\frac{{\partial}#2}{{\partial}#1}}}

\newcommand{\pint}{-\!\!\!\!\!\!\!\int}

\newcommand{\op}[1]{\hat{#1}}

\newcommand{\av}[1]{\left\langle\, #1\,\right\rangle}

\newcommand{\br}[1]{\langle\,#1\,|\,}
\newcommand{\ke}[1]{\,|\,#1\,\rangle}

\newcommand{\Tr}[1]{{\rm Tr}\left\{ #1\right\}}

\newcommand{\de}{{\!\rm d}}

\newcommand{\e}{{\rm e}}

\newcommand{\g}{{\!\,=\,\!}}
\newcommand{\n}{\vec{\nabla}}

\newcommand{\ii}{{\rm i}}

\newcommand{\sa}{\left[ \begin{array} {c} }
\newcommand{\se}{\end{array}\right]}

\begin{document}
\title{Lorentz atom revisited by solving Abraham--Lorentz equation of motion}
\author{J. Bosse}
\affiliation{Fachbereich Physik, Freie Universit{\"a}t Berlin, 14195 Berlin, Germany}
\date{\today}
\begin{abstract}
By solving the non--relativistic Abraham-Lorentz (AL) equation, I demonstrate that AL equation of motion is {\em not} suited for treating the Lorentz atom, because a steady--state solution does not exist. The AL equation serves as a tool, however, for deducing appropriate parameters $\Omega,\Gamma$  to be used with the {\em equation of forced oscillations} in modelling the Lorentz atom.
The electric polarizability, which many authors ``derived'' from AL equation in recent years, is found to violate Kramers--Kronig relations rendering obsolete the extracted photon--absorption rate, for example. Fortunately, errors turn out to be small quantitatively, as long as light frequency $\omega$ is neither {\em too close to} nor {\em too far from} resonance frequency $\Omega$.
Polarizability and absorption cross section are derived for the Lorentz atom by purely classical reasoning and shown to agree with quantum--mechanical calculations of the same quantities. In particular, oscillator parameters $\Omega,\,\Gamma$ deduced by treating the atom as a quantum oscillator are found equivalent to those derived from classical AL equation. The instructive comparison provides a deep insight into understanding the great success of Lorentz's model which was suggested long before the advent of quantum theory.

\end{abstract}
\maketitle

\section{Introduction}

In recent years, the classical {\sc Lorentz}--oscillator model serving as an intuitive description of an atom under the influence of the $ac$--electric field associated with a standing wave of visible light has celebrated a revival in quantum--optics literature \cite{gwo:00}, \cite{gaf:10}. While the classical {\em equation of forced oscillations}, with a friction force proportional to the $1^{\rm st}$ time--derivative of elongation, has many applications (e.\,g., the $ac$--current circuit with impedance and capacity \cite{bes.b:61} or simplified models of density fluctuations in liquids \cite{mar:68}) besides the {\sc Lorentz} atom, the latter has played a special role. When deriving  the so--called `radiative reaction force'  from classical electrodynamics to account for energy loss of an accelerated charge by radiation, {\sc Abraham} and {\sc Lorentz} (AL) arrived at a modified equation of motion with a friction force proportional to the $3^{\rm rd}$ time--derivative of the oscillator elongation. The radiative reaction force has been discussed extensively for more than 100 years, both for relativistic and non--relativistic velocities of the charged particle, because its implications have raised fundamental problems such as ``pre--acceleration'' or ``run--aways in the absence of external forces'' which are still open for discussion  (see, e.\,g., \cite[Ch.\,17]{jac:62}, \cite{mur:77}, \cite{jic:87}, \cite[Ch.\,11]{gri:99}). In case of the {\sc Lorentz} atom, the oscillating electron may be described non--relativistically. So the non--relativistic AL equation, only, will be studied here.

In Sec.\,\ref{Sec-oscillator-model},  a concise review is presented of the unique solution of the {\em forced--oscillations equation}, inclusive of its steady--state limit, by introducing {\em classical} elongation response and relaxation functions.

In Sec.\,\ref{Sec-Lorentz-atom}, the  unique solution of the AL equation for given initial values $(x_0,\,v_0,\,b_0)$ is determined. The unique solution is shown to be a ``run--away'' implying non--existence of a steady--state solution and spotting the AL equation of motion as an inappropriate tool for describing the {\sc Lorentz} atom. The {\em forced--oscillation equation}  is suggested, instead, to do the job together with oscillator parameters $(\Omega,\,\Gamma)$ derived from AL equation. In this context, a widely spread error is pointed out regarding the ``complex polarizability'' of the {\sc Lorentz} atom (see, e.\,g., \cite[Sec.\,17.9]{jac:62}, \cite[Sec.\,II.A]{gwo:00}, \cite[Sec.\,2A]{gaf:10}).

In Sec.\,\ref{Sec-QM}, for a quantum--mechanical system perturbed by an oscillatory external field, a representation--free perturbation expansion in the field strength is presented for an expectation value. With its help, the  `absorbed power' (dipole moment to $1^{\rm st}$ order) and the `$ac$--{\sc Stark} shift' (energy to $2^{\rm nd}$ order) are derived in terms of the  dipole--dipole response function, or rather the complex polarizability.
The {\em quantum--mechanical} response function is evaluated for a charged quantum oscillator and compared to the {\em  classical} dipole--response function of the Lorentz atom derived in Sec.\,\ref{Sec-Lorentz-atom}. Perfect agreement between classical and quantum--mechanical calculation is found, which renders an explanation for the great success of the classical Lorentz atom.

In Sec.\,\ref{Sec-Conclusions}, the reader finds a Summary and Conclusions. In Sec.\,\ref{Sec-Appendix}, Appendix, the unique solution of the AL equation of motion for given initial values ($x_0,\,v_0,\,b_0$) is presented in terms of classical response and relaxation functions. In addition, the Appendix contains a short compendium on integral transforms used in this work.

\section{Classical oscillator}
\label{Sec-oscillator-model}
\subsection{Response and relaxation functions}
The ordinary second--order differential equation
\1
\label{forced-osc-eq}
\ddot{x}(t)+\Gamma\dot{x}(t)+\Omega^2 x(t)=f(t)/m
\2
with positive constants ($m,\,\Omega,\,\Gamma$) and external force $f(t)$ has for given initial values,
\1
\label{osc-initial-values}
x_0= x(t_0)\;,~~~~~~~~~~~~~~~~~v_0=\dot{x}(t_0)\;,
\2
a unique solution. Finding this solution belongs to the first exercises in every math course on ordinary differential equations.  For physical applications, it is useful to cast the unique solution into the intuitive form ($t\ge t_0$),
\bea
\label{forced-osc-eq-sol}
x(t;t_0)&=&\phi(t-t_0)x_0+\ii\chi(t-t_0) m v_0\\
&&\hspace{4em}+\int_{t_0}^t\de t'~\ii\chi(t-t')f(t')\;,\nonumber
\eea
with abbreviations
\1
\label{osc-resp-relax}
\chi(t)=\frac{\e^{\zeta_1 t}-e^{\zeta_2 t}}{\ii m (\zeta_1-\zeta_2)}\;,~~~~~~
\phi(t)=\frac{\zeta_1\e^{\zeta_2 t}-\zeta_2\e^{\zeta_1 t}}{\zeta_1-\zeta_2}~~~~~(t\ge0)
\2
denoting, resp., {\em classical} elongation--response and  (normalized) --relaxation function defined, at this stage, for non--negative arguments, only.
Here $\zeta_1$ and $\zeta_2$ denote roots of the characteristic polynomial associated with Eq.\,(\ref{forced-osc-eq}),
$\zeta_{1,\,2}=-\Gamma/2\pm\sqrt{(\Gamma/2)^2-\Omega^2}$, which obey
\1
\Re[\zeta_{1,\,2}]<0\;,~~~~~~~\zeta_1\zeta_2=\Omega^2\;,~~~\zeta_1+\zeta_2=-\Gamma\;.
\2
Due to {\em negative} real parts of both roots $\zeta_1$ and $\zeta_2$, the functions $\phi(t)$ and $\chi(t)$ will decay to zero if time arguments grow large.

It is convenient to extend the definitions of response and relaxation functions to {\em negative} time arguments. In accordance with quantum--mechanical linear--response theory (see Sec.\,\ref{Sec-QM} below), I postulate
\1
\label{symmetries}
\chi(-t)=-\chi(t)=\chi(t)^*\;,~~~~~\phi(-t)=\phi(t)=\phi(t)^*\;.
\2
After introducing phase angle $\vartheta$ and frequency $\tilde{\Omega}$ by
\1
\label{def-om-tilde}
\vartheta=\arctan\left[\frac{\Gamma}{2\tilde{\Omega}}\right],~~\tilde{\Omega}=\left\{
\begin{array}{cc}
\sqrt{\Omega^2-(\Gamma/2)^2}&,~\Omega>\Gamma/2\\
\ii\sqrt{(\Gamma/2)^2-\Omega^2}&,~\Omega\le\Gamma/2
\end{array}
\right.\;,
\2
which will be real valued, if $\Gamma<2\Omega$ (low--damping regime), the response and relaxation functions may be expressed in the more descriptive way,
\1
\label{trig-osc-resp-relax}
\chi(t)=\e^{-\frac{\Gamma}{2}|t|}\,\frac{\sin(\tilde{\Omega}t)}{\ii m \tilde{\Omega}}\;,~~\phi(t)=\e^{-\frac{\Gamma}{2}|t|}\,\frac{\cos(\tilde{\Omega}|t|-\vartheta)}{\cos\vartheta}\;.
\2
Here occurrence of $|t|$ reflects the symmetry introduced in Eq.\,(\ref{symmetries}).

\subsection{Steady--state elongation}
\label{Sec-steady-state}

The initial time $t_0$ in Eqs.\,(\ref{osc-initial-values}),(\ref{forced-osc-eq-sol}) is properly interpreted as the instant, when the external force $f(t)$ is switched on. After switch--on, the elongation $x(t,t_0)$ will at first depend on $t_0$ and initial values ($x_0,\,v_0$) until `transients' have died off due to relaxation processes, and the system described by Eq.\,(\ref{forced-osc-eq}) acquires a steady state. The corresponding {\em steady--state elongation} $\xi(t)$ is found  by switching on the force $f(t)$ adiabatically and choosing $t_0=-\infty$ in Eq.\,(\ref{forced-osc-eq-sol}),
\1
\label{osc-steady-state-sol}
\xi(t)=\lim_{t_0\to-\infty}x(t,t_0)=\int_0^\infty\de t'~\e^{-o t'}\ii\chi(t')f(t-t')\;.
\2
Adiabatical switch--on is described by replacing  under the integral in Eq.\,(\ref{forced-osc-eq-sol}): $f(t')\to f(t')\e^{-o(t-t')}$ ($o>0$).  Subsequent substitution $t-t'\to t'$ results in Eq.\,(\ref{osc-steady-state-sol}).  It is understood  from here on that $o\to0$ is taken {\em after} time integrations have been performed ---without repeatedly employing the explicit notation $\lim_{o\to0}$.   This convention regarding treatment of the small positive frequency $o$ will be used throughout.

It is to be noted that the steady--state elongation Eq.\,(\ref{osc-steady-state-sol}) is {\em independent} of initial values $(x_0,\,v_0)$, because the general solution of the {\em homogeneous} equation (Eq.\,(\ref{forced-osc-eq}) for $f(t)\!\equiv\!0$),
$
x^{\rm h}(t,t_0)=\phi(t-t_0)x_0+\ii\chi(t-t_0) m v_0
$
which {\em does} depend on initial conditions, will vanish in the steady--state limit. This independence of initial values is a physical requirement on a {\em steady--state} solution, because initial values $x_0$ and $v_0$ are not (and cannot usually be) measured.

\subsection{Dynamical susceptibility}

If the force entering the integrand in Eq.\,(\ref{osc-steady-state-sol}) is represented by its {\sc Fourier} integral, the steady--state solution will also appear in {\sc Fourier}--expanded form,
\1
\label{FT-osc-steady-state}
\xi(t)=\frac{1}{2\pi}\int_{-\infty}^\infty\de\omega~\xi_\omega\exp(-\ii t\omega)\;,~~~~~~\xi_\omega=\tilde{\chi}(\omega+\ii o) f_\omega\;,
\2
with the {\em  dynamical susceptibility} $\tilde{\chi}(\omega+\ii o)= \xi_\omega/f_\omega$ determining the ratio between {Fourier}--transformed elongation and force. Here $\tilde{\chi}(z)$, the {\sc Fourier--Laplace} transform (FLT, see Sec.\ref{Sec-FLT} for details) of the response function $\chi(t)$,  has been introduced.
For the classical response and relaxation functions given in Eq.\,(\ref{trig-osc-resp-relax}), FLTs are readily calculated ($s={\rm sign}\Im[z]$),
\1
\label{chi-of-z}
\tilde{\chi}(z)=\frac{1/m}{\Omega^2-z(z+s \ii \Gamma)}\;,~~~
\tilde{\phi}(z)=\frac{z+s \ii \Gamma}{\Omega^2-z(z+s \ii \Gamma)}\;.
\2
For the dynamical susceptibility, one has
\1
\label{dyn-susc-1}
\tilde{\chi}(\omega+\ii o)=\frac{\Omega^2}{\Omega^2-\omega^2-\ii\omega\Gamma}~\tilde{\chi}_0\equiv\chi'(\omega)+\ii\chi''(\omega)
\2
with real and imaginary parts
\bea
\label{chi-react}
\chi'(\omega)&=&\frac{(\Omega^2-\omega^2)\Omega^2}{(\Omega^2-\omega^2)^2+(\omega\Gamma)^2}~\tilde{\chi}_0\;,\\
\label{chi-dissip}
\chi''(\omega)&=&\frac{\omega\Gamma\Omega^2}{(\Omega^2-\omega^2)^2+(\omega\Gamma)^2}~\tilde{\chi}_0\;,
\eea
and the {\em static susceptibility}
\1
\tilde{\chi}_0= \frac{1}{m\Omega^2}=\tilde{\chi}(\ii o)=\chi'(0)
\;.
\2
It is worth pointing out the close relationship between response and relaxation function known as {\sc Kubo} identity,
\1
\label{KI}
\chi(z)\equiv\frac{z\tilde{\phi}(z)+1}{m\Omega^2}~~\Longleftrightarrow~~\chi(t)\equiv\frac{\ii\dot{\phi}(t)}{m\Omega^2}\;,
\2
which is reflected by Eqs.\,(\ref{trig-osc-resp-relax}),\,(\ref{chi-of-z}), and also
mentioning the  exact rewriting,
\1
\label{dyn-susc-2}
\tilde{\chi}(z)=\frac{\Omega}{\tilde{\Omega}}\left[\frac{\frac{\Gamma}{2}}{\tilde{\Omega}-\left(z+s\ii\frac{\Gamma}{2}\right)}
+\frac{\frac{\Gamma}{2}}{\tilde{\Omega}+\left(z+s\ii\frac{\Gamma}{2}\right)}\right]\frac{\Omega\tilde{\chi}_0}{\Gamma}\;,
\2
which highlights the resonance patterns  emerging near $\omega=\pm\tilde{\Omega}$ (cf.\,Eq.\,(\ref{def-om-tilde})) in case of  $\Gamma/\Omega\ll1$.

Finally, it is important to notice that the two ingredients $\chi'(\omega)=\chi'(-\omega)$  and $\chi''(\omega)=-\chi''(-\omega)$  of the dynamical susceptibility are intimately connected via {\sc Kramers--Kronig} relations, cf.\,Eq.\,(\ref{Kramers-Kronig}) below.
These {\em dispersion relations} are an immediate consequence of the generalized susceptibility $\tilde{\chi}(z)$ appearing as the {\sc Fourier--Laplace} transform of the response function $\chi(t)$. Violation of {\sc Kramers--Kronig} relations is an indicator for a faulty determination of $\tilde{\chi}(\omega+\ii o)$. Similarly, experimental results on $\chi'(\omega)$ and $\chi''(\omega)$ would not be trustworthy, if available measured data permitted someone to demonstrate violation of {\sc Kramers--Kronig} relations.

\subsection{Oscillatory force}

The steady--state solution Eq.\,(\ref{osc-steady-state-sol}) acquires a specially simple form, if one assumes a sinusoidal $t$--dependence of frequency $\omega$ for the force,
\1
\label{osc-force}
f(t)=f_0\cos(\omega t)\equiv f_0\Re\left[\e^{\mp\ii t\omega}\right]\;.
\2
with real $f_0$ and $\omega$. Inserting this force into Eq.\,(\ref{osc-steady-state-sol}), results in the {\em steady--state elongation}
\bea
\xi(t)&=&\Re\left[\tilde{\chi}(\omega+\ii o)f_0\e^{-\ii t\omega}\right]\nonumber\\
\label{xi-of-t}
&=&\left[\chi'(\omega) \cos(\omega t)+\chi''(\omega)  \sin(\omega t)\right] f_0
\eea
which may be cast into the clearly arranged form
\1
\xi(t)=A\,\cos\left(\omega t-\varphi\right)
\2
with $\omega$--dependent amplitude and phase shift,
\bea
A&=&\frac{\Omega^2\tilde{\chi}_0 f_0}{\sqrt{(\Omega^2-\omega^2)^2+(\omega\Gamma)^2}}\le A_{\rm m} =\frac{\Omega^2\tilde{\chi}_0 f_0}{\tilde{\Omega}\,\Gamma}\;,\\
\varphi&=&\arctan\left(\frac{\omega\Gamma}{\Omega^2-\omega^2}\right)+\pi\frac{1-{\rm sign}\left(\Omega^2-\omega^2\right)}{2}\;.\nonumber
\eea
The oscillator  picks up energy from the oscillatory force and {\em dissipates} this energy via friction ($\Gamma\!>\!0$). The work done by the external force during time interval $(t,\, t +\de t)$  amounts to  $[\xi(t+\de t)- \xi(t)] f(t) = \de t~ \dot{\xi}(t)f(t)$. Integrating
this energy over an oscillation period $T= 2\pi/\omega$ and dividing by $T$ results in
the {\em average absorbed power}
\1
\label{absorbed-power}
P(\omega)=\frac{1}{T}\int_0^T\de t~ \dot{\xi}(t)f(t) =\frac{1}{2}\,\omega\chi''(\omega)\,f_0^2 ~\ge0\;.
\2
A glance at Eq.\,(\ref{chi-dissip}) shows that no power will be absorbed, if  a constant external force is applied ($\omega=0$), while maximum power absorption will be achieved, if  the `resonance value'  $\omega_{\rm r}=\pm\Omega$ is chosen for the applied oscillatory--force frequency $\omega$.
Finally, it is to be noted that the driven elongation $\xi(t)$ develops its amplitude maximum $A_{\rm m}$ for a different driving--field frequency $\omega_{\rm m}=\pm\Omega\sqrt{1-\Gamma^2/(2\Omega^2)}$. Moreover, both $\omega_{\rm r}$ and $\omega_{\rm m}$ differ  from $\tilde{\Omega}$ in Eq.\,(\ref{trig-osc-resp-relax}). For $\Gamma\ll\Omega$, however, the three frequencies $|\omega_{\rm m}|<\tilde{\Omega}<|\omega_{\rm r}|$ will differ only slightly, and merge for $\Gamma\to0$.

\section{Lorentz atom}
\label{Sec-Lorentz-atom}

\subsection{Abraham--Lorentz equation}

The forced--oscillations Eq.\,(\ref{forced-osc-eq})  presents an ingenious model first suggested by {\sc Lorentz} for describing an atom under the influence of visible light. {\sc Lorentz} assumed an electron (charge $q=-e$, mass $m= m_{\rm e}$) which is bound to the atomic nucleus by a restoring force $f_\Omega(t)=-m\Omega^2 x(t)$ and subject to a friction force $f_\Gamma(t)=-m\Gamma\dot{x}(t)$. If light is shining on the atom this electron will, in addition, be exposed  to an oscillating force $f(t)= q E_0\cos(\omega t)$ exerted on a charge by the electric field associated with a standing light wave of frequency $\omega$ (neglecting much smaller magnetic--field contributions).
While the value of the restoring--force parameter $\Omega^2$ was roughly known, because $\Omega\approx10^{15}\,{\rm s}^{-1}$ could be detected by finding the light--wave frequency $\omega_{\rm r}$ `in resonance' with the atom, there was little experimental information on the extremely small but finite damping constant ($\Gamma\ll\Omega$) at the end of the nineteenth century. In summary, the {\sc Lorentz}--model parameters $\Omega$ and $\Gamma$ had to be determined from theoretical reasoning.

In the {\sc Abraham--Lorentz} (AL) equation of motion \cite[Eq.\,(17.9)]{jac:62},
\1
\label{AL-eq}
\ddot{x}(t)-\tau\dddot{x}(t)+\omega_0^2 x(t)=f(t)/m\;,
\2
the radiation--reaction force $f_{\rm RR}(t)= m\tau \dddot{x}(t)$ replaces the unknown friction force $f_{\Gamma}(t)=-m\Gamma\dot{x}(t)$ of the forced--oscillator equation of motion (Eq.\,(\ref{forced-osc-eq})), while the resonance frequency which determines the restoring force has been denoted by $\omega_0$ here, for clarity reasons.
The radiation--reaction force has been derived from classical electrodynamics. It accounts for the energy loss which the {\em accelerated} electron will suffer due to {\sc Hertz} radiation of electromagnetic waves.  The parameter
\1
\label{tau-classic}
\tau=2R/(3c)= q^2/(6\pi\epsilon_0\,m c^3)
\2
with classical charge radius $R= q^2/(4\pi\epsilon_0 mc^2)$, permittivity $\epsilon_0$, and light velocity $c$ in vacuum denotes a very short characteristic time. For the time it ``takes light to pass by an electron'', one finds $\tau\approx10^{-23}\,{\rm s}$ resulting in the small parameter $\tau\omega_0\approx10^{-8}$ for an atomic electron.

In view of the smallness of the characteristic time $\tau$ and the dimensionless parameter $(\tau\omega_0)$, it is tempting to rewrite the AL equation
\bea
\ddot{x}&=&-\omega_0^2 x+\frac{f}{m}+\tau\dnd{t}{}\ddot{x}\nonumber\\
&=&\left(1-\tau\dnd{t}{}\right)^{-1}\left[-\omega_0^2 x+\frac{f}{m}\right]\nonumber\\
&=&\left[1+\tau\dnd{t}{}+{\cal O}(\tau^2)\right]\left[-\omega_0^2 x+\frac{f}{m}\right]\nonumber\\
\label{AL-eq-approx}
&=&-\omega_0^2x-\tau\omega_0^2\dot{x}+\frac{f(t+\tau)}{m}+{\cal O}(\tau^2)\;.
\eea
In the representation Eq.\,(\ref{AL-eq-approx}),  one of AL equation's  strange properties shows up: the acceleration at (present time) $t$, $\ddot{x}(t)$, is induced by a force $f(t+\tau)$ to be applied at (future time) $t+\tau$. Leaving aside philosophical questions arising from the `pre--acceleration' problem (see, e.g., \cite[Sec.\,17.7]{jac:62}, \cite[Sec.\,11.2.2]{gri:99}) and, in view of $\tau\omega_0\ll1$, simply replacing   $f(t+\tau)\to f(t)$ by assuming a sufficiently slowly varying force,
the expansion Eq.\,(\ref{AL-eq-approx}) shows that  the widely used  {\sc Abraham--Lorentz} equation (Eq.\,(\ref{AL-eq})) could be replaced to a very good approximation with the {\em equation of forced oscillations} (Eq.\,(\ref{forced-osc-eq})), if damping constant  and restoring--force constant were chosen as follows: $\Gamma\to\tau\omega_0^2$ and $\Omega^2\to\omega_0^2$.

Instead of further endeavours to find an approximate equation by exploiting the smallness of $\tau$,  let us solve the AL equation of motion itself.

\subsection{Roots of AL characteristic polynomial}
\label{Sec-Root-AL}

The inhomogeneous third--order ordinary differential equation with constant coefficients may be solved by `brute force'.
It is straight--forward to find the {\em unique} solution of Eq.\,(\ref{AL-eq}) for given initial conditions
\1
\label{AL-initial-1}
x_0=x(t_0)\;,~~~~ v_0= \dot{x}(t_0)\;,~~~~~b_0=\ddot{x}(t_0)\;.
\2
The unique solution $x_{\rm AL}(t,t_0)= x_{\rm AL}^{\rm h}(t,t_0) +x_{\rm AL}^{\rm p}(t,t_0)$, which is the sum of the general solution of the {\em homogeneous} and one {\em particular solution} of the inhomogeneous equation, will then be used to derive the steady--state elongation $\xi_{\rm AL}(t)= x_{\rm AL}(t,t_0\to-\infty)$ following the procedure applied in Sec.\,\ref{Sec-steady-state}.

\begin{figure}
\begin{center}
\includegraphics[height=50mm,width=70mm]{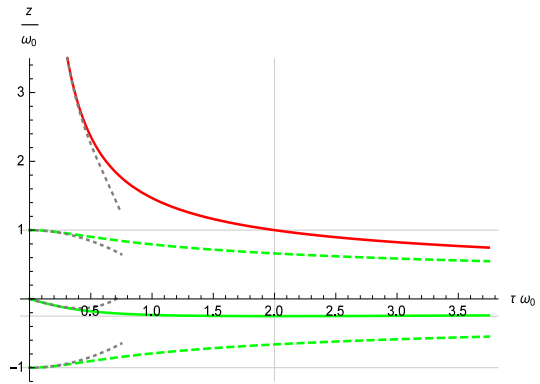}
\caption{
\label{AL-roots-plot}
Roots of characteristic polynomial. Solid red: $\zeta_2=1/\tau-2u=\omega_0[(\tau\omega_0)^{-1}+(\tau\omega_0)-2(\tau\omega_0)^3+\dots]>0$.
Solid green: $\Re[\zeta_{1,3}]=-\omega_0[(\tau\omega_0)/2-(\tau\omega_0)^3+\dots]<0$. Dashed green: $\Im[\zeta_1]=\omega_0[1-5(\tau\omega_0)^2/8+\dots]>0$, $\Im[\zeta_3]=-\Im[\zeta_1]$. Dashed gray: small--$(\tau\omega_0)$ asymptotes. Grid lines mark  $\Im[\zeta_{1,2}]$ extrema at $(\tau\omega_0=0,v/\omega_0=\pm1)$, and $\Re[\zeta_{1,3}]$ minimum at $(\tau\omega_0=2,\,u/\omega_0=-1/4)$.
}
\end{center}
\end{figure}
Denoting by $\zeta_1,\,\zeta_2,\,\zeta_3$ the roots of the characteristic polynomial  associated with Eq.\,(\ref{AL-eq}),
\1
\zeta^2-\tau \zeta^3+\omega_0^2 =0\;,
\2
one finds, as expected for the roots of a $3^{\rm rd}$ order polynomial, a pair of complex--conjugate besides a real root,
\1
\label{AL-roots-1}
\zeta_1=u+\ii v\;,~~~~~\zeta_2=\frac{1}{\tau}-2u\;,~~~~~\zeta_3=u-\ii v
\2
with  real and imaginary parts of $\zeta_1$ given by
\bea
\label{AL-roots-2}
u&=&-\frac{\left(w-1\right)^2}{6\tau w}\le0\;,~~~~~~~~v=\frac{1-w^2}{2\tau w\sqrt{3}}\ge0\;,\nonumber\\
w&=&\left[1+\frac{3}{2}\tau\omega_0\left(9\tau\omega_0-\sqrt{12+81\tau^2\omega_0^2}\right)\right]^{\frac{1}{3}}\;,
\eea
where $0\le w\le1$. Real and imaginary parts of the characteristic--polynomial roots in Eq.\,(\ref{AL-roots-1}) are displayed in Fig.\,\ref{AL-roots-plot} as function of the parameter $\tau\omega_0$.

\subsection{Absence of AL steady--state solution}
\label{Sec-no-steady-state}

It is important to realize that the {\em positive} root $\zeta_2$ (red line in Fig.\,\ref{AL-roots-plot}) implies a unique solution $x_{\rm AL}(t,t_0)$ which will diverge in the {\em steady--state} limit,
\1
\label{xi-AL-steady-state}
\xi_{\rm AL}(t)=\lim_{t_0\to-\infty}x_{\rm AL}(t,t_0)=\infty\;,
\2
for generic initial conditions (see Appendix for details). Evidently, a steady--state solution of the AL equation does {\em not}, in general, exist.
Facing this staggering fact, one must admit that Eq.\,(\ref{AL-eq}) is {\em not} suited to model the steady--state elongation of a {\em bounded} atomic electron.
This conclusion is corroborated by closer inspection of the special case, $f(t)\equiv0$:
\1
\label{AL-eq-hom}
\ddot{x}(t)-\tau\dddot{x}(t)+\omega_0^2 x(t)=0\;.
\2
Introducing abbreviations
\1
\label{Gamma-Omega}
\Gamma=-2u\;,~~~~~~~\Omega^2=u^2+v^2\;,
\2
the general solution of the {\em homogeneous} AL equation, Eq.\,(\ref{AL-eq-hom}), may be cast into the form ($t\ge t_0$)
\bea
\label{AL-sol-hom}
x_{\rm AL}^{\rm h}(t,t_0)&=&\phi(t-t_0)x_0+\ii\chi(t-t_0) m v_0\\
&&+ \e^{(t-t_0)\zeta_2}\left[b_0+\Gamma v_0+\Omega^2 x_0\right]t_1^2(t,t_0)\nonumber\;,
\eea
which explicitly shows the contribution that will diverge, due to $\zeta_2>0$, when $(t-t_0)$ grows large. That holds, because $t_1^2(t,t_0)$ abbreviates an expression, which reduces to the positive constant $t_1^2(t,-\infty)=\tau^2/(1+4v^2\tau^2)$  in the steady--state limit, while $\phi$ and $\chi$ denote relaxation and response function, resp., defined in Eq.\,(\ref{osc-resp-relax}) (or Eq.\,(\ref{trig-osc-resp-relax}), equivalently) ---for parameters ($\Gamma$, $\Omega^2$) provided in Eq.\,(\ref{Gamma-Omega}).
Hence, the  first line on r.h.s. of Eq.\,(\ref{AL-sol-hom}), which evidently represents the general homogeneous--equation solution of Eq.\,(\ref{forced-osc-eq}), will vanish in the steady--state limit.

An interesting aspect of the `run--away solution' Eq.\,(\ref{AL-sol-hom}) is the observation that the diverging contribution would have been absent, if one assumed initial values {\em not} chosen independently as in Eq.\,(\ref{AL-initial-1}) but  in such a way that the pre--factor of $\e^{(t-t_0)\zeta_2}$ in brackets on the r.h.s. of Eq.\,(\ref{AL-sol-hom}) will vanish,
\1
\label{AL-initial-2}
\ddot{x}(t_0)+\Gamma \dot{x}(t_0)+\Omega^2 x(t_0)=0\;,
\2
with $\Gamma=\Gamma(\tau,\omega_0)$ and $\Omega^2=\Omega^2(\tau,\omega_0)$ given in Eq.\,(\ref{Gamma-Omega}).
This condition will prevent the solution of Eq.\,(\ref{AL-eq-hom}) from running away, ---a noteworthy observation, because {\em any} instant of time could have been chosen to play the role of initial time $t_0$.
I conclude:
\begin{enumerate}
\item
Equation\,(\ref{AL-initial-2}) must hold  at any time $t$, if  {\em bounded elongations} of the {\sc Lorentz}--atom electron are to be guaranteed.
\item
\label{conclusion-2}
The {\em equation of forced oscillations} (Eq.\,(\ref{forced-osc-eq}) and properties discussed in Sec.\,\ref{Sec-oscillator-model})  with parameters
\bea
\label{Gamma-Omega-expanded}
\Gamma&=&-2u\nonumber\\&=&\tau\omega_0^2\left[1-2(\tau\omega_0)^2+{\cal O}\left((\tau\omega_0)^4\right)\right]\;,\nonumber\\
\Omega&=&\sqrt{u^2+v^2}\nonumber\\&=&\omega_0\left[1-(\tau\omega_0)^2/2+{\cal O}\left((\tau\omega_0)^4\right)\right]
\eea
should be used for treating the {\sc Lorentz} atom.
\item
My conclusions are corroborated by the fact that the suggested procedure is consistent with the small--$\tau$ expansion given in Eq.\,(\ref{AL-eq-approx}).
\end{enumerate}

\subsection{Lorentz--atom polarizability}
\label{Sec-Lorentz-at-polariz}

Following the conclusion of Sec.\,\ref{Sec-no-steady-state}\,-\,item\,\ref{conclusion-2}, the  atomic dipole moment $d(t)=(-e)\xi(t)$, which is induced by the electric field of a standing light wave  exerting  the force $f(t)= (-e) E_0\cos(\omega t)$ on the electron (within dipole approximation), can be read from Eq.\,(\ref{xi-of-t}),
\1
\label{ind-dipol-moment}
d(t)=\Re[\tilde{\chi}(\omega+\ii o)e^2\, E_0\e^{-\ii t\omega}]
\2
with oscillator parameters
\1
\Omega\to\omega_0\;,~~~~~~~~~\Gamma\to\tau\omega_0^2
\2
taken from Eq.\,(\ref{Gamma-Omega-expanded}) ---with perfectly sufficient precision in consideration of $\tau\omega_0\approx10^{-8}$.
The {\sc Lorentz} model also allows to account for additional damping processes besides radiative loss by replacing $\Gamma$ in Eq.\,(\ref{ind-dipol-moment})  with a {\em total} damping constant $\Gamma_{\rm t}$,
\1
\label{Gamma-total}
\Gamma\to\Gamma_{\rm t}=\Gamma+\Gamma'\;,
\2
which, as opposed to \cite[Eq.\,(17.61)]{jac:62}, does {\em not} depend on frequency.

In view of the constant dipole moment $d(t)= d_0$ induced by a static field $E_0$,
\1
\label{static-ind-dipol}
d_0=(\tilde{\chi}_0e^2)\,E_0 =\alpha E_0~~~~~~~~~(\omega=0)
\2
with $\alpha= \tilde{\chi}_0e^2$ denoting the (static) {\em  polarizability}, it has become common to name the dynamical dipole susceptibility, $\tilde{\alpha}(\omega+\ii o)= \tilde{\chi}(\omega+\ii o)e^2$, the ``complex polarizability'' \cite[Sec.\,II.A]{gwo:00}. Its real part, the (generalized $\omega$--dependent) polarizability
\1
\label{alpha-of-omega}
\alpha(\omega)=\Re[\tilde{\alpha}(\omega+\ii o)]=\chi'(\omega)e^2\;,
\2
determines a force ${\bf F}\g-\n U_{\rm dip}$ acting on the atom in the light field, where
\1
U_{\rm dip}({\bf r})=-\frac{1}{2}\alpha(\omega)\overline{|{\bf E}({\bf r},t)|^2}=-\chi'(\omega)e^2\frac{E_0^2}{4}
\2
denotes the `optical dipole potential' which will be identified  as the average atomic energy shift, known as `ac--{\sc Stark} effect' in Sec.\,\ref{Sec-ac-Stark-effect} below. Within classical electrodynamics, the optical dipole potential can only be made plausible to within a factor 2, because one has $-\overline{{\bf d}(t)\cdot{\bf E}_0\cos(\omega t)}=-\chi'(\omega)e^2E_0^2/2$ for
the time--averaged potential energy of an electric dipole moment in an external electric field.

\subsection{Absorption and scattering of radiation by Lorentz atom}

The imaginary part of the dynamical dipole susceptibility, $\Im[\tilde{\alpha}(\omega+\ii o)]=\chi''(\omega)e^2$,  via Eq.\,(\ref{absorbed-power}) determines the average power $P(\omega)$ absorbed by the atom from the electric field, which implies the absorption cross--section ($\lambdabar_0=c/\omega_0$, wavelength at resonance, divided by $2\pi$),
\1
\label{sigma-abs}
\sigma_{\rm abs}(\omega)=\frac{P(\omega)}{\epsilon_0 c E_0^2/2}=6\pi\lambdabar_0^2\frac{\Gamma\Gamma_{\rm t}\omega^2}{\left(\omega_0^2-\omega^2\right)^2+\left(\omega\Gamma_{\rm t}\right)^2}
\2
which obeys the famous $f$--sum rule,
\1
\label{f-sum-rule}
\int_0^\infty\de\omega~\sigma_{\rm abs}(\omega)~=\frac{\pi }{2 \epsilon_0  c}\frac{e^2}{m}\;,
\2
also known as `dipole sum rule' in the present context. It must be emphasized that the $f$-sum rule, valid for both classical and quantum mechanical systems, states the following interesting fact. The  integrated absorption cross section on the r.h.s. of Eq.\,(\ref{f-sum-rule}) is determined by the ratio $e^2/m$ alone. It does not depend on further details of the system, here represented by oscillator frequency and damping constants.

A photon-absorption rate $\Gamma_{\rm abs}(\omega)$ has been considered in \cite[Sec.\,II.A]{gwo:00} which is determined by $\chi''(\omega)$, too.  From quantum--mechanical scattering theory, one finds ($\Theta(x)$ denoting unit--step function)
\1
\label{Gamma-sc}
\Gamma_{\rm abs}(\omega)=\frac{\chi''(\omega)}{2\hbar}\,e^2 E_0^2~\Theta(\omega)=\frac{P(\omega)}{\hbar\omega}~\Theta(\omega)\;,
\2
if the atom is assumed in its electronic ground state (i.\,e., at zero temperature) when hit by photons. In Eq.\,(\ref{Gamma-sc}), $\Gamma_{\rm abs}(-\omega)=0$ for $\omega>0$ expresses the fact that an atom in its ground state cannot loose energy  by stimulated (or spontaneous) emission of a photon of energy $\hbar\omega$. It can only win energy  by absorbing a photon of energy $\hbar\omega$.

Finally, the time--dependent dipole moment induced by the oscillatory external field (Eq.\,(\ref{ind-dipol-moment})) will produce an electromagnetic field which in the far--field dipole--approximation $({\bf E},{\bf B})_{\rm rad}=\frac{\ddot{d}(t-r/c)}{4\pi\epsilon_0c^2 r}\left[{\bf e}\times\frac{{\bf r}}{r}\right]\left(\times\frac{{\bf r}}{r},\,\frac{1}{c}\right)$  may be interpreted in terms of a radiation--scattering cross section \cite[Eq.\,(2A.48)]{gaf:10},\,\cite[Eq.\,(17.63)]{jac:62}
\bea
\label{resonant-Lorentz-sc}
\sigma_{\rm sc}(\omega)&=&\frac{8\pi}{3}R^2\frac{\omega^4}{\left(\Omega^2-\omega^2\right)^2+\left(\omega\Gamma_{\rm t}\right)^2}\\
&\to&6\pi\lambdabar_0^2\left(\frac{\Gamma}{\omega_0}\right)^2\left\{\begin{array}{cl}
\omega^4/\omega_0^4&,\,\omega\ll\omega_0\\
\frac{(\omega_0/2)^2}{(\omega-\omega_0)^2+(\Gamma_{\rm t}/2)^2}&,\,\omega\approx\omega_0\\
1&,\,\omega\gg\omega_0
\end{array}
\right.\nonumber
\eea
from which well--known scattering regimes  are easily identified as limiting cases: {\sc Rayleigh} scattering for $\omega\ll\Omega$, {\sc Thomson} scattering for $\omega\gg\Omega$, and, for $\omega\approx\Omega$, the resonant {\sc Lorentz} scattering exhibiting the characteristic line shape with `full width at half maximum (FWHM)', $\Gamma_{\rm t}$, and `peak cross section',
\1
\sigma_{\rm sc}(\Omega)=\frac{8\pi}{3}R^2\frac{\Omega^2}{\Gamma_{\rm t}^2}=6\pi\lambdabar_0^2\,\left(\frac{\Gamma}{\Gamma_{\rm t}}\right)^2\;.
\2
Here $R=3 c\tau/2$ denotes the classical electron radius, and oscillator parameters are given by $\Omega=\omega_0$ and total decay constant $\Gamma_{\rm t}=\Gamma+\Gamma'$ with $\Gamma=\tau\omega_0^2$.

As opposed to the statement in \cite[Eq.\,(17.72)]{jac:62} which refers to {\em all} $\omega$, Eqs.\,(\ref{sigma-abs}) and (\ref{resonant-Lorentz-sc}) imply $\sigma^{\rm L}_{\rm abs}(\omega)=\sigma^{\rm L}_{\rm sc}(\omega)+\sigma^{\rm L}_{\rm r}(\omega)$ for frequencies $|\omega-\omega_0|\ll\omega_0$ only, i.\,e., for the resonant {\sc Lorentz}--absorption (or total) cross section. The total cross section  is composed of a scattering contribution $\sigma^{\rm L}_{\rm sc}(\omega)$, spelled out  in Eq.\,(\ref{resonant-Lorentz-sc}) ($\omega\approx\omega_0$), and a `reaction cross section' $\sigma^{\rm L}_{\rm r}(\omega)$ with the same {\sc Lorentz} resonance denominator, but $\Gamma$ replaced with $\sqrt{\Gamma\Gamma'}$ in the numerator. Consequently, $\sigma^{\rm L}_{\rm abs}(\omega)$ must be given by $\sigma^{\rm L}_{\rm sc}(\omega)$ with $\Gamma$ replaced by $\sqrt{\Gamma\Gamma_{\rm t}}$ in the numerator, which is easily verified from Eq.\,(\ref{sigma-abs}). The reason for the discrepancy with Ref.\,\cite{jac:62} will become clear in Sec.\,\ref{Sec-Pitfalls}.

\subsection{Pitfalls}
\label{Sec-Pitfalls}

Regarding the classical model of the atomic complex polarizability, much confusion has been created  in literature by erroneous conclusions drawn from the AL equation of motion,  Eq.\,(\ref{AL-eq}), with oscillatory external force $f(t)= f_0 \cos(\omega t)$.
In  Refs.\,\cite{jac:62}, \cite{gri:99}, \cite{gwo:00}, \cite{gaf:10}, e.\,g., and in numerous other publications, authors search for a {\em particular} solution of Eq.\,(\ref{AL-eq}) which oscillates with frequency $\omega$ of the driving force. Indeed, there is one such solution,
\1
\label{x-osc}
x^{\rm osc}(t)=\frac{(\omega_0^2-\omega^2)\cos(\omega t)+\tau\omega^3 \sin(\omega t)}{(\omega_0^2-\omega^2)^2+(\tau\omega^3)^2}\frac{f_0}{m}\;,
\2
which can be checked easily by inserting $x^{\rm osc}(t)$ into Eq.\,(\ref{AL-eq}). But $x^{\rm osc}(t)$ is {\em not} the steady--state solution of the inhomogeneous AL equation.  According to Eq.\,(\ref{xi-AL-steady-state}), such a steady--state solution does not exist which brought me to rule out the AL equation of motion as a candidate for describing the {\sc Lorentz}--atom elongation.

It must be emphasized that, in contrast to my findings, Eq.\,(\ref{x-osc}) is frequently claimed to present the steady--state solution to the AL equation with oscillatory force, which is not true as I demonstrated in Sec.\,\ref{Sec-no-steady-state}.
Since $x^{\rm osc}(t)$ is {\em not} the steady--state solution, we are {\em not} allowed to interpret Eq.\,(\ref{x-osc}) as if it were the analog of Eq.\,(\ref{xi-of-t}).
Extracting from Eq.\,(\ref{x-osc}) a ``susceptibility'',
\1
\label{X-of-omega}
X(\omega)=\frac{\omega_0^2}{\omega_0^2-\omega^2-\ii \tau\omega^3}X_0\;,~~~~~~~X_0=\frac{1}{m\omega_0^2}
\2
is a frequently repeated mistake which
\begin{itemize}
\item
is found already in the high--impact monograph \cite{jac:62}, where in \cite[Eqs.\,(17.60-61)]{jac:62} a non--radiative decay constant $\Gamma'$ was assumed in addition to the radiative decay constant $\Gamma=\tau\omega_0^2$, both of which were combined into a {\em total decay constant} $\Gamma_{\rm t}(\omega)=\Gamma'+(\omega/\omega_0)^2 \Gamma$, which is evidently not constant! Moreover, $\Gamma_{\rm t}(\omega)$ violates the $f$--sum rule and suppresses the high--frequency Thomson scattering in \cite[Eq.\,(17.63)]{jac:62}.
According to my findings in Eqs.\,(\ref{Gamma-Omega-expanded}--\ref{ind-dipol-moment}), the total decay rate must here read $\Gamma_{\rm t}=\Gamma'+\Gamma$ as arrived at in Eq.\,(\ref{Gamma-total}) above, which will repair the mentioned deficiencies.
\item
was made in the monograph \cite[Ex. 11.4, p.\,468]{gri:99} implicitly, when claiming $\Gamma=\tau\omega^2$ instead of the correct result Eq.\,(\ref{Gamma-Omega-expanded}),
\item
has been carried on into the optical--dipole potential community by the very informative and often cited review article \cite[Sec. II.\,A]{gwo:00},
\item
is even found in the more recent monograph \cite{gaf:10}, where it shows up in \cite[Eq.\,(2A.53)]{gaf:10} and again,  as a nasty suppressor of {\sc Thomson} scattering, in \cite[Eq.\,(2A.48)]{gaf:10}.
\item
would also result, if one erroneously applied to Eq.\,(\ref{AL-eq}) the {\em mnemonic trick}  which is so helpful in remembering $\tilde{\chi}(\omega+\ii o)$.

Namely, {\sc Fourier} trans\-forming  Eq.\,(\ref{forced-osc-eq}) which is, of course, obeyed  by the steady--state elongation $\xi(t)$,
\bea
\label{short-cut}
m\left[(-\ii\omega)^2+(-\ii\omega)\Gamma+\Omega^2\right]\xi_\omega=f_\omega\nonumber\\
=\left[\tilde{\chi}(\omega+\ii o)\right]^{-1}\xi_\omega\;,
\eea
and reading the result Eq.\,(\ref{dyn-susc-1}) for $\tilde{\chi}(\omega+\ii o)$  from the {\sc Fourier}--transformed equation of motion.

Note that the ``short--cut'' Eq.\,(\ref{short-cut}) works out alright only, because I proved in Sec.\,\ref{Sec-steady-state} above that Eq.\,(\ref{forced-osc-eq})  does indeed have a unique steady--state solution. The same, however, does not hold true for the AL equation (Eq.\,(\ref{AL-eq})) as I demonstrated in Sec.\,\ref{Sec-no-steady-state} above.
\end{itemize}

\begin{figure}
\begin{center}
\includegraphics[height=50mm,width=70mm]{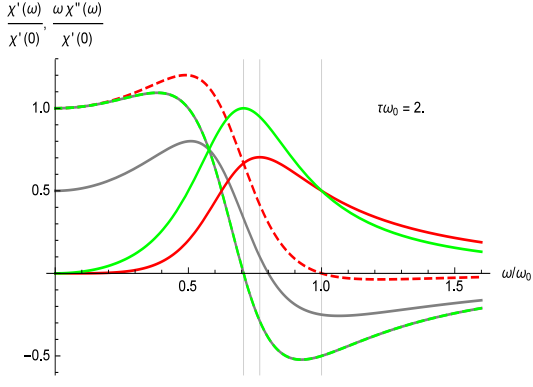}
\caption{
\label{KKcheck2E-00}
Kramers--Kronig check. Full green: $\omega\chi''(\omega)/\tilde{\chi}_0$ (Eq.\,(\ref{chi-dissip})); dashed green: $\chi'(\omega)/\tilde{\chi}_0$ (Eq.\,(\ref{chi-react})); full red: $\omega\Im[X(\omega)]/X_0$ (Eq.\,(\ref{X-of-omega})); dashed red: $\Re[X(\omega)]/X_0$ (Eq.\,(\ref{X-of-omega})); full gray: integrals on r.h.s. (Eq.\,(\ref{KK-check-on-X}) \& first Eq.\,(\ref{Kramers-Kronig})). Grid lines indicate positions of $\Omega$, $\omega_{\rm m}^{X}$, $\omega_0$ (from left to right).
}
\end{center}
\end{figure}
But why can $X(\omega)$ not serve as a proper dynamical susceptibility, anyway? Answer: Because it does {\em not} obey {\sc Kramers--Kronig} relations,
\1
\label{KK-check-on-X}
\Re[X(\omega)]\ne\frac{1}{\pi}\pint_{-\infty}^\infty\de\bar{\omega}~\frac{\Im[X(\bar{\omega})]}{\bar{\omega}-\omega}\;,
\2
which is a consequence of $x^{\rm osc}(t)$ not being the steady--state solution of the AL equation.
Inequality Eq.\,(\ref{KK-check-on-X}) is clearly demonstrated in Fig.\,\ref{KKcheck2E-00}, where the full gray line (cutting the ordinate at $\approx0.5$) displays the numerically evaluated principal--value integral from r.h.s. of Eq.\,(\ref{KK-check-on-X}) which has been divided by the constant $X_0$. This should be compared with $\Re[X(\omega)]/X_0$ depicted as dashed red line. Both curves differ markedly indicating  violation of the {\sc Kramers--Kronig} relation. As pointed out above, however, a proper susceptibility must obey this relation.
As opposed to $X(\omega)$, the real and imaginary parts of the dynamical susceptibility in Eq.\,(\ref{dyn-susc-1}) do form a {\sc Kramers--Kronig} pair. This has also been demonstrated in Fig.\,\ref{KKcheck2E-00}: the gray line representing the numerically evaluated principal--value integral (first Eq.\,(\ref{Kramers-Kronig}) for $f''(\omega)\to\chi''(\omega)$, after dividing by $\tilde{\chi}_0$) cannot be distinguished from the dashed green line displaying $\chi'(\omega)/\tilde{\chi}_0$ from Eq.\,(\ref{chi-react}).
The very large parameter value chosen  in Fig.\,\ref{KKcheck2E-00} for demonstration purposes, $\tau\omega_0=2$, requires exact evaluation of $\Gamma$ and $\Omega^2$ using Eq.\,(\ref{Gamma-Omega}) with Eq.\,(\ref{AL-roots-2}).

The difference between correct (green) and faulty (red) model polarizability curves will diminish for decreasing values of $\tau\omega_0$. This observation is substantiated by the relations
\bea
\label{X-chi-relations-1}
\frac{\Re[\tilde{X}(\omega)]}{\chi'(\omega)}&=&1-(\tau\omega_0)^2\left\{1-\frac{\omega^4\left[1+{\cal O}\left((\tau\omega_0)^2\right)\right]}{(\omega_0^2-\omega^2)\omega_0^2}\right\}\nonumber\\
\\
\label{X-chi-relations-2}
\frac{\Im[\tilde{X}(\omega)]}{\chi''(\omega)}&=&\frac{\omega^2}{\omega_0^2}~\left\{1-
(\tau\omega_0)^2\frac{\omega^2\left[1+{\cal O}\left((\tau\omega_0)^2\right)\right]}{\omega_0^2}\right\}\nonumber\\
\eea
found from comparison of Eq.\,(\ref{X-of-omega}) with  Eqs.\,(\ref{chi-react}--\ref{chi-dissip}), where $\Gamma$ and $\Omega$ are given in Eq.\,(\ref{Gamma-Omega-expanded}). For electron parameters ($\tau\omega_0\approx10^{-8}$), it seems that use of the {\em incorrect}  $\tilde{X}(\omega)$  will produce {\em quantitatively} acceptable polarizability results, if one restricts to the frequency range
\1
\label{range-omega}
1\gg\left|1-\frac{\omega^2}{\omega_0^2}\right|\gg(\tau\omega_0)^2
\2
which allows to set $\omega^2/\omega_0^2\approx1$ in Eq.\,(\ref{X-chi-relations-2}) and, at the same time, keep sufficient distance to the pole in Eq.\,(\ref{X-chi-relations-1}).
\begin{figure}
\begin{center}
\includegraphics[height=50mm,width=70mm]{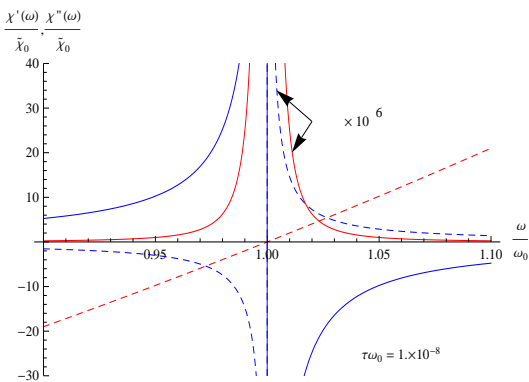}
\caption{
\label{chi-re-im}
Polarizability: reactive (blue) and dissipative (red, $10^6$--fold amplified) parts. Errors from Eqs.\,(\ref{X-chi-relations-1}),(\ref{X-chi-relations-2}): $\{\Re[X(\omega)]/\chi'(\omega)-1\}\times10^2$ (blue--dashed, $10^6$--fold amplified) and  $\{\Im[X(\omega)]/\chi''(\omega)-1\}\times10^2$ (red--dashed) reflecting factor $\omega^2/\omega_0^2$ in Eq.\,(\ref{X-chi-relations-2}).
}
\end{center}
\end{figure}
The limitations imposed on the range of frequencies by Eq.\,(\ref{range-omega}) are best appreciated by throwing a glance at Fig.\,\ref{chi-re-im}, where relative errors derived from Eqs.\,(\ref{X-chi-relations-1}--\ref{X-chi-relations-2}) are displayed together with $\chi'(\omega)$ and $\chi''(\omega)$ for the realistic value $\tau\omega_0=10^{-8}$ referring to the oscillating electron of the {\sc Lorentz} atom. From Fig.\,\ref{chi-re-im} it is clear that  big quantitative errors ($\pm$20\% for $|\omega/\omega_0-1|\approx0.1$) only occur in the imaginary part and in a detuning range, where $\chi''(\omega)$ is very small, while the quantitative error in the real part is negligibly small in magnitude ($<3.\times 10^{-5}$ \% for $|\omega/\omega_0-1|>0.005$).

On the one hand, this observation may possibly explain, why the analytically incorrect polarizability $X(\omega)$ (Eq.\,(\ref{X-of-omega})) could survive in literature for so long without being debunked.

On the other hand, Fig.\,\ref{chi-re-im} clearly demonstrates that the photon--absorption rate $\Gamma_{\rm abs}(\omega)$ deduced from Eq.\,(\ref{X-of-omega})  is systematically under-- (over--)estimating the true photon--absorption rate in the red-- (blue--)detuned frequency regime (see, e.g., in \cite[Eqs.\,(11),(41)]{gwo:00}).
For the experimentally relevant ratio of absorption rate and optical dipole potential (cf.\,\cite[Eq.\,(14)]{gwo:00}), one finds from Eqs.\,(\ref{Gamma-sc}),(\ref{absorbed-power}) the simple result,
\1
\label{ratio-Gamma-U}
\frac{\hbar\Gamma_{\rm abs}}{U_{\rm dip}}=\frac{2\omega \Gamma}{\omega^2-\Omega^2}
=\left\{\begin{array}{rl}
-2\omega\Gamma/\Omega^2 & \left[1+{\cal O}\left(\omega^2/\Omega^2\right)\right]\\
\Gamma/\Delta & \left[1+{\cal O}\left(\Delta/\Omega\right)\right]
\end{array}
\right.
\2
where $\Delta=\omega-\Omega$ and $\omega\ge0$.
For a ``quasi--electrostatic trap'' (QUEST), this ratio  is under--estimated by roughly $100$\% (for $\omega\ll\Omega$), because $\Gamma_{\rm abs}$ will be larger  by the factor $(\Omega/\omega)^2$ if deduced from  Eq.\,(\ref{X-of-omega}). For a ``far off--resonance trap'' (FORT), which is  understood to be detuned sufficiently slightly to still obey $|\Delta|\ll\Omega$, the incorrect polarizability happens to produce the correct ratio $\Gamma/\Delta$ given in Eq.\,(\ref{ratio-Gamma-U}), because $\omega\approx\Omega$.

\section{Quantum--mechanical reasoning}
\label{Sec-QM}

\subsection{Perturbation expansion for expectation value}
\label{Sec-pert-expansion}

A physical system (Hamiltonian $\hat{H}$) will take on the explicitly time--dependent Hamiltonian
\1
\label{H_t}
\hat{H}_t=\hat{H}-\hat{D} E_0 \cos(\omega t)
\2
under the influence of an external, linearly polarized, oscillatory field ${\bf e}E_0\cos(\omega t)$ which couples to the dynamical variable $\hat{D}={\bf \hat{D}\cdot e}$. As a consequence, the average value at time $t$ of an arbitrary operator $\hat{O}_t$ will deviate from its stationary--state value,
\bea
\label{av-O-t}
\av{\delta \hat{O}_t}_t&\equiv&\av{\hat{O}_t}_t-\av{\hat{O}_t}\nonumber\\
&=&\av{\hat{O}_t}_t^{(1)}+\av{\hat{O}_t}_t^{(2)}~+~\dots\;.
\eea
Assuming the perturbing field switched on {\em adiabatically} at time $t_0=-\infty$ (which amounts to replacing $E_0\cos(\omega t')\to\e^{-o(t-t')}E_0\cos(\omega t')$ with $o>0$ for all $t'\le t$), the $n$--th order in $E_0$ contribution  to $\av{\delta \hat{O}_t}_t$ reads explicitly ($\tau_0=0$),
\bea
\label{av-O-t-nth-order}
\av{\hat{O}_t}_t^{(n)}&=&\ii^n\int_{\tau_0}^\infty\de\tau_1\dots\int_{\tau_{n-1}}^\infty\de\tau_n~
\chi_{\hat{O}_t^\dagger;\hat{D}}(\tau_1,\dots,\tau_n)
\nonumber\\
&&\hspace{3em}\times\prod_{j\g1}^n\e^{-o\tau_j}E_0\cos[\omega(t-\tau_j)]
\eea
with the $n$--th order response function,
\bea
\label{chi-nth-order}
&&\chi_{\hat{A};\hat{B}}(\tau_1,\dots,\tau_n)=\\&&
\frac{1}{\hbar^n}\av{\comm{\dots\comm{   \comm{\hat{A}^\dagger}{\hat{B}(-\tau_1)}  }{\hat{B}(-\tau_2)} \dots}{\hat{B}(-\tau_n)}}.\nonumber
\eea
Here $\hat{B}(t)=\exp(\ii t\hat{H}/\hbar)\hat{B}\exp(-\ii t\hat{H}/\hbar)$ denotes a {\sc Heisenberg} operator  referring to the unperturbed system, and the stationary--state average is defined as
\1
\av{\hat{A}}=\Tr{\hat{A} \hat{W}}\;, ~~~~\comm{\hat{H}}{\hat{W}}=0
\2
with statistical operator $\hat{W}$ describing the initial stationary state of the unperturbed system.

It must be emphasized that $\av{\delta \hat{O}_t}_t$ in Eqs.\,(\ref{av-O-t}--\ref{chi-nth-order}) describes the {\em steady--state} deviation from the unperturbed expectation value, which is induced by the external field. Interestingly enough, the first--order result $\av{\hat{O}_t}_t^{(1)}$ (written out explicitly  in Eq.\,(\ref{qm-induced-dipole}) for the induced dipole moment) has the same structure found for the  steady--state solution in Eq.\,(\ref{osc-steady-state-sol}) for the {\em classical} oscillator elongation.

\subsection{Induced dipole moment}

For the atomic dipole moment induced by a linearly--polarized standing light wave of frequency $\omega$, one reads for the first--order result $d(t)=\av{\delta \hat{D}}_t^{(1)}$ from Eq.\,(\ref{av-O-t-nth-order}),
\bea
\label{qm-induced-dipole}
d(t)&=&\ii\int_0^{\infty}\de\tau~\e^{-o\tau}\chi_{\hat{D};\hat{D}}(\tau)E_0\cos[\omega(t-\tau)]\nonumber\\
&=&\Re\left[\tilde{\chi}_{\hat{D};\hat{D}}(\omega+\ii o) E_0\e^{-\ii\omega t}\right]\;,
\eea
with the dipole--dipole response function
\1
\label{def-dipole-resp-fn}
\chi_{\hat{D};\hat{D}}(t)=\frac{1}{\hbar}\av{\comm{\hat{D}(t)}{\hat{D}}}=-\chi_{\hat{D};\hat{D}}(-t)
\;,
\2
where $\hat{D}$ is identified with the component in field direction of the atomic dipole--moment operator ${\bf \hat{D}}$.

The corresponding dynamical dipole--susceptibility (``complex polarizability'') resulting from {\sc Fourier--Laplace} transforming $\chi_{\hat{D};\hat{D}}(t)$ according to  Eq.\,(\ref{def-FLT}), may quite generally be cast into the form \cite{mar:68}
\bea
\label{dipole-susc-general}
\tilde{\chi}_{\hat{D};\hat{D}}(z)&=&\frac{\Omega_D^2}{\Omega_D^2-z^2-z \tilde{K}_D(z)}\,\tilde{\chi}_{\hat{D};\hat{D}}(\ii o)\;.
\eea
This formally exact expression is cited here only to point out the following facts.
\begin{itemize}
\item
The relaxation kernel
$\tilde{K}_D(z)=\int_{-\infty}^\infty\frac{\de\overline{\omega}}{\pi}\frac{K_{D}''(\overline{\omega})}{\overline{\omega}-z}$ is determined by an even, non--negative, and bounded spectral function $K_{D}''(\omega)$. This generally frequency--dependent `total damping constant' $K_{D}''(\omega)=\Im[\tilde{K}_{D}(\omega+\ii o)]$ will inevitably be associated with a resonance--frequency renormalization via $\Re [\tilde{K}_{D}(\omega+\ii o)]$. Such a real contribution is missing in \cite[Eq.\,(17.60)]{jac:62} resulting in the violation of {\sc Kramers--Kronig} relations and $f$--sum rule discussed in Sec.\,\ref{Sec-Pitfalls} above. Moreover, $\Gamma_{\rm t}(\omega)$ in \cite[Eq.\,(17.61)]{jac:62} is not bounded.
\item
The relaxation kernel $\tilde{K}_D(z)$  is the FLT of the dipole memory function $K_D(t)$ governing the generalized oscillator equation,
\1
\label{gen-osc-eom}
\ddot{\phi}_{\hat{D};\hat{D}}(t)+\Omega_D^2\phi_{\hat{D};\hat{D}}(t)+\int_0^t\de t'K_D(t')\dot{\phi}_{\hat{D};\hat{D}}(t-t')=0
\2
with initial conditions $\phi_{\hat{D};\hat{D}}(0)=1$, $\dot{\phi}_{\hat{D};\hat{D}}(0)=0$, which is obeyed by the (normalized) dipole relaxation function. Both, Eq.\,(\ref{dipole-susc-general}) and Eq.\,(\ref{gen-osc-eom}), are  formally exact and, in view of {\sc Kubo}'s identity Eq.\,(\ref{KI}),  equivalent statements.
\end{itemize}
To conclude these general remarks, I emphasize that memory effects may  be neglected in some applications, rendering
\1
K_D(t)\approx2\Gamma\delta(t)\Longrightarrow K_{D}''(\omega)\approx\Gamma\Longrightarrow \tilde{K}_D(z)\approx s\ii\Gamma
\2
a reasonable approximation ---as is the case for the quantum oscillator in Sec.\,\ref{Sec-qm-oscillator}. Under these circumstances, Eq.\,(\ref{gen-osc-eom}) reduces to a {\em free--oscillations equation} of the same type obeyed by the classical relaxation function $\phi(t)$ introduced in Sec.\,\ref{Sec-oscillator-model}.  These remarks on very general quantum mechanical (and quantum statistical)  results may illuminate the great success of models such as the {\sc Lorentz} atom, which are based on the classical {\em forced--oscillations equation} of motion.

\subsection{Quantum oscillator}
\label{Sec-qm-oscillator}

Assuming the eigenvalue problem of the unperturbed Hamiltonian solved ($\op{H}\ke{n}=\ke{n}\varepsilon_n$, $n=0,1,2,\dots$) and the atom in its ground state initially ($\op{W}=\ke{0}\br{0}$), the dipole--response function defined in Eq.\,(\ref{def-dipole-resp-fn}) is easily evaluated,
\1
\label{dipole-resp-fn-damped}
\chi_{\hat{D};\hat{D}}(t)=\frac{2}{\ii\hbar}\sum_{n\ne0}\left|D_{n0}\right|^2\sin(\omega_{n0}t)~\e^{-\frac{\Gamma_n}{2}|t|}\;,
\2
with the  dynamical susceptibility,
\1
\label{dipole-susc-damped}
\tilde{\chi}_{\hat{D};\hat{D}}(z)=\sum_{n\ne0}\frac{2m\omega_{n0}}{\hbar e^2}\left|D_{n0}\right|^2
\frac{\Omega_n^2\,\tilde{\chi}_n(\ii o)}{\Omega_n^2-z^2-s\ii\Gamma_n z}\;,
\2
given by {\sc Fourier--Laplace} transformation.
Here $D_{n0}=\br{n}{\bf \hat{D}}\cdot{\bf e}\ke{0}$ are dipole--moment matrix elements and $\omega_{n0}=(\varepsilon_n-\varepsilon_0)/\hbar$  denote atomic excitation frequencies ($n\g1,2,\dots$). Abbreviations have been introduced for partial static polarizabilities and resonance frequencies,  $\tilde{\chi}_n(\ii o)=e^2/(m\Omega_n^2)$ and $\Omega_n^2=\omega_{n0}^2+(\Gamma_n/2)^2$, respectively.

In Eq.\,(\ref{dipole-resp-fn-damped}), ad--hoc damping factors have been inserted which account approximately for the {\em natural lifetimes} of excited atomic states while  preserving the symmetry spelled out in Eq.\,(\ref{def-dipole-resp-fn}).
Excited atomic states are well known to have a finite natural lifetime $\tau_n=1/\Gamma_n$  even if no electromagnetic field is applied, because there is ``spontaneous emission'' due to the atom interacting with vacuum fluctuations, interactions which have not been included into the unperturbed Hamiltonian $H$.  In leading order (electric dipole transitions), spontaneous emission will occur at a rate \cite[Chap. V]{hei:53}
\1
\label{atomic-line-width-1}
\Gamma_n=\frac{4\alpha}{3c^2}\sum_{n'}^{\varepsilon_{n'}<\varepsilon_{n}}\omega_{nn'}^3\left|\br{n'}{\bf \op{r}}\ke{n}\right|^2\;,
\2
where $\alpha=e^2/(4\pi\epsilon_0~\hbar c)\approx1/137$ denotes the {\sc Sommerfeld} fine--structure constant.

It is very instructive to evaluate the dipole--response function in detail for a simple model of an atom.
Within the quantum--oscillator model for the atomic electron, $\hat{H}=\hbar\omega_{10}(\hat{a}^\dagger\hat{a}+1/2)$, one has for the electric dipole--moment operator $\hat{D}=-e x_0(\hat{a}^\dagger+\hat{a})$ with oscillator length $x_0=\sqrt{\hbar/(2m_{\rm e}\omega_{10})}$ resulting in matrix elements
$\left|D_{n0}\right|^2=e^2x_0^2~\delta_{n,1}$, which leave only a single term in the sum on r.h.s of Eq.\,(\ref{dipole-resp-fn-damped}),
\1
\label{chi-quantum-osc}
\chi_{\hat{D};\hat{D}}(t)= e^2\frac{\sin(\omega_{10} t)}{\ii m_{\rm e}\omega_{10}}~\e^{-\frac{\Gamma_1}{2}|t|}\;.
\2
Evaluation of the damping constant $\Gamma_1$ using the transition rate  of Eq.\,(\ref{atomic-line-width-1}) results in
\1
\label{quant-class-identif}
\begin{array}{cccl}
\omega_{10}&=&(E_1-E_0)/\hbar&\longleftrightarrow~\tilde{\Omega}\\
\Gamma_1&=&\bar{\tau}\omega_{10}^2&\longleftrightarrow~\Gamma
\end{array}\;,
\2
where the characteristic time $\bar{\tau}$ turns out to be identical to the time constant $\tau$ introduced in Eq.\,(\ref{tau-classic}),
\1
\bar{\tau}\equiv\tau=\frac{2\alpha\hbar}{3 m_{\rm e}c^2}\approx6.3\times10^{-24}\,{\rm s}\;.
\2
The quantum--mechanical results derived above are noteworthy in several respects, as they demonstrate why the {\em classical  oscillator} model discussed in Sec.\,\ref{Sec-oscillator-model} has been so extremely successful in describing an atom irradiated by light.
\begin{enumerate}
\item
The general result Eq.\,(\ref{qm-induced-dipole}) for the induced dipole moment of any physical system in a weak electric field has the same formal structure as one finds for the steady--state elongation of a classical oscillator subjected to an external field, see Eq.\,(\ref{osc-steady-state-sol}).
\item
\label{item-chiDD}
Dipole--dipole response function of a quantum oscillator (Eq.\,(\ref{chi-quantum-osc})) and  elongation--response function of a classical oscillator (Eq.\,(\ref{trig-osc-resp-relax})) become identical ---after multiplying the latter by $(-e)^2$ and identifying induced moment $(-e)x(t)\to\av{\delta\hat{D}}_t$,  force $f(t)\to (-e)E(t)$, and
fixing oscillation frequency  and damping constant of the classical {\sc Lorentz} atom according to Eq.\,(\ref{quant-class-identif}).
\end{enumerate}
Note that the latter identification solves, by quantum--mechanical arguments, the problem   of finding appropriate parameters $\Omega,\Gamma$ to be used for the classical Lorentz--atom: $\Omega=\omega_{10}\left[1+{\cal O}\left((\tau\omega_{10})^2\right)\right]$, $\Gamma=\tau\omega_{10}^2$, which to leading order in the small parameter $(\tau\omega_{10})$ agree with the classical solution, provided one also identifies $\hbar$ times the classical resonance frequency  $\omega_0$ with the energy difference $(E_1-E_0)$, i.\,e.,  $\omega_0\equiv\omega_{10}$.

{\sc Lorentz} and {\sc Abraham}  at the end of 19$^{\rm th}$ century, of course, did not have recourse to results from quantum theory (cf.\,Eqs.\,(\ref{chi-quantum-osc}--\ref{quant-class-identif})). They had to specify their model parameters by using classical electrodynamics, only.
While  $\Omega$ in Eq.\,(\ref{forced-osc-eq}) could naturally be associated with the frequency of resonantly absorbed light, determination of $\Gamma$ required introduction of a {\em radiative reaction force} which lead to the strange new AL equation of motion  (Eq.\,(\ref{AL-eq})) for the oscillator elongation.

It is therefore noteworthy and comforting to see that the {\em classical} radiation--damping constant $\Gamma$ derived from AL equation (Eq.\,(\ref{Gamma-Omega-expanded})) is perfectly reproduced by the quantum--mechanical result in Eq.\,(\ref{quant-class-identif}).

\subsection{Average absorbed power}
\label{Sec-abs-power}

The average power absorbed from the external oscillatory field by the physical system described in Eq.\,(\ref{H_t}) is given by
\bea
\label{absorbed-power-qm}
P(\omega)&=&\overline{\der{t}{\av{H_t}_t}}=\overline{\av{\pd{t}{H_t}}_t}=\overline{\av{\delta \hat{D}}_t\sin(\omega t)}E_0\nonumber\\
&=&\frac{1}{2}\,\omega \chi''_{\hat{D};\hat{D}}(\omega)\,E_0^2~+~{\cal O}(E_0^3)\;,
\eea
where $\overline{F(t)}=\frac{1}{T}\int_0^T\de t~F(t)$, $T=2\pi/\omega$ and, in the last line, use has been made of Eq.\,(\ref{qm-induced-dipole}).
The quantum--mechanical result in lowest non--vanishing order of perturbation theory, Eq.\,(\ref{absorbed-power-qm}), should be compared to the classical expression Eq.\,(\ref{absorbed-power}). As in case of the induced dipole moment, the formal structures of both, quantum and classical, results for $P(\omega)$ are identical. The average power absorbed from the $ac$-electric field by a charged quantum oscillator in its ground state will coincide with the power absorbed by the classical oscillator, because of equivalent response functions, cf.\,Sec.\,\ref{Sec-qm-oscillator}\,-\,item\,\ref{item-chiDD}.

\subsection{$ac$--{\sc Stark} effect and optical dipole potential}
\label{Sec-ac-Stark-effect}

The energy $\av{H}$ of an atom is expected to change upon applying an electric field $E_0\cos(\omega t)$. Such a phenomenon is well known as {\sc Stark} shift in case of a constant electric field ($\omega=0$). Since an atom in its ground state has no permanent dipole moment, the {\sc Stark} shift is typically of $2^{\rm nd}$ order in $E_0$.   The rapidly oscillating electric field of visible light will also induce a shift of the atomic energy, which is rapidly   oscillating with frequency $\omega$ and known as $ac$--{\sc Stark} effect.  Due to the high frequency of light, the induced shift cannot be detected by time--resolved measurements. Therefore, only the time--averaged  shift is of interest here (averaging over period $T=2\pi/\omega$).

Applying the perturbation expansion Eq.\,(\ref{av-O-t}) to the operator of total energy, $\hat{O}_t\to \hat{H}_t$, the averaged induced energy shift is
\bea
\overline{\av{\delta \hat{H}_t}_t}&=&\overline{\av{\hat{H}_t}_t}-\av{H}=\Delta\varepsilon+{\cal O}(E_0^3)\;,\nonumber\\
\label{ac-shift-1}
\Delta\varepsilon&=&-E_0\overline{\av{\hat{D}}_t^{(1)}\cos(\omega t)}+\overline{\av{\hat{H}}_t^{(2)}}\;.
\eea
Since one may replace  under the time average $\av{\hat{D}}_t^{(1)}\to\av{\delta\hat{D}}_t^{(1)}$, the first contribution to $\Delta \varepsilon$ is easily evaluated with the help of the induced dipole moment in Eq.\,(\ref{qm-induced-dipole}),
\1
\label{ac-shift-2}
-E_0\overline{\av{\hat{D}}_t^{(1)}\cos(\omega t)}=-\chi'_{\hat{D};\hat{D}}(\omega)\frac{E_0^2}{2}\;.
\2
Here $\chi'_{\hat{D};\hat{D}}(\omega)=\alpha(\omega)$ is the electric polarizability defined quantum--mechanically, which should be compared to its  classical pendent in Eq.\,(\ref{alpha-of-omega}).
The second contribution to $\Delta \varepsilon$ (Eq.\,(\ref{ac-shift-1})) is read from Eq.\,(\ref{av-O-t-nth-order}). Noting the relation
\1
\label{ac-shift-3}
\chi_{\hat{H};\hat{D}}(\tau_1,\tau_2)=\frac{1}{\ii}\pd{\tau_2}{}\chi_{\hat{D};\hat{D}}(\tau_2-\tau_1)
\2
between quadratic and linear dipole--dipole response function, and employing $\sin(\omega t)/t\to\pi\delta(t)$ for large $\omega$ under the final integral, one finds
\bea
\overline{\av{\hat{H}}_t^{(2)}}&=&\ii^2\int_0^\infty\de\tau_1\int_{\tau_1}^\infty\de\tau_2~\chi_{\hat{H};\hat{D}}(\tau_1,\tau_2)\nonumber\\
&&\times
\e^{-o(\tau_1+\tau_2)}\cos[\omega(\tau_2-\tau_1)]\frac{E_0^2}{2}\nonumber\\
&=&\frac{\ii}{2}\int_0^\infty\de t~\e^{-o t}\chi_{\hat{D};\hat{D}}(t)\nonumber\\
&&\times \left[\cos(\omega t)+\frac{\omega}{o}\sin(\omega t)\right]\frac{E_0^2}{2}\nonumber\\
\label{ac-shift-4}
&=&\frac{1}{2}\chi'_{\hat{D};\hat{D}}(\omega)\frac{E_0^2}{2}\;,
\eea
which is just $(-1/2)$ times the energy of the induced dipole moment in the external field.
Summing both contributions, Eq.\,(\ref{ac-shift-2}) and Eq.\,(\ref{ac-shift-4}), there will be a non--zero average shift of the system energy induced by the electric field (``$ac$--{\sc Stark} effect''),
\bea
\label{ac-Stark-shift}
\Delta\varepsilon=-\frac{1}{4}\chi'_{\hat{D};\hat{D}}(\omega)E_0^2\;.\\
\nonumber
\eea
As expected, the conventional quadratic {\sc Stark} shift follows from Eq.\,(\ref{ac-Stark-shift}) for $\omega=0$.
If the external electric field is produced, e.\,g., by the standing wave of linearly (in $z$--direction) polarized  light created by two laser beams counter--propagating along $x$--axis,
\[
E_0\cos(\omega t)=\tilde{E}_0\cos({\bf k}\cdot{\bf r}-\omega t)+\tilde{E}_0\cos(-{\bf k}\cdot{\bf r}-\omega t)\;,
\]
the field strength  $E_0\to2\tilde{E}_0\cos(2\pi x/\lambda)$  will acquire a spatial dependence, $E_0=E_0({\bf r})$. As long as field variations over distances of the order of system diameter are negligible, which is the case for an atom in visible light ($\lambda\gg a_0$), Eq.\,(\ref{ac-Stark-shift}) applies. The energy shift ---and thus the energy of the atom itself, too---  will be a function of the atomic position ${\bf r}$ via $E_0({\bf r})^2$ resulting in a force acting on the atom (``dipole force'', $-\n \Delta\varepsilon({\bf r})$). Hence one defines an ``optical--dipole potential'',
\1
U_{\rm dip}({\bf r})=\Delta\varepsilon({\bf r})\;,
\2
which crucially depends on the frequency of the laser light used to produce the potential via the electric polarizability $\alpha(\omega)=\chi'_{\hat{D};\hat{D}}(\omega)$.

\section{Conclusions}
\label{Sec-Conclusions}

By determining the unique solution of the non--relativistic AL equation (Eq.\,(\ref{AL-eq})), which turns out to be a `run--away' for generic initial conditions, I showed that there is no steady--state solution that will describe the driven oscillations of an atomic dipole moment induced by the electric field of light. Due to its run--away solution, Eq.\,(\ref{AL-eq}) does not qualify for modelling the bounded electron of {\sc Lorentz}'s atom.

Therefore, an attempt to determine the complex atomic polarizability by employing any one particular solution of the AL equation, which is {\em not} the steady state, will be a misleading effort.  The erroneous ``polarizability'' Eq.\,(\ref{X-of-omega}) which, besides other deficiencies, violates {\sc Kramers--Kronig} relations and $f$--sum rule, has spread widely in literature. The error is obviously invoked by (and has been traced back to) authors' unjustified assumption of having found the steady--state solution of the AL equation which, as I proved by finding the unique solution Eq.\,(\ref{AL-unique-sol-1}), does not exist.

However, according to the discussion in Sec.\,\ref{Sec-no-steady-state}, there is also a positive aspect of the AL equation. In an endeavour to account for radiative dissipation processes within classical electrodynamics, the  AL equation  allows to determine the appropriate oscillator parameters $\Omega$ and $\Gamma$ to be used with Eq.\,(\ref{forced-osc-eq}),  when implementing radiative dissipation in the {\sc Lorentz} atom.

Finally, in Sec.\,\ref{Sec-QM} the {\em steady--state} induced dipole moment of a system placed into an external electric field is studied by quantum mechanical perturbation theory in a `semi--classical approach'. The quantum mechanical dipole--dipole response function, which determines electric polarizability, average power absorbed from the field, and optical dipole potential, is identified as a {\em quantum} analog of the {\em classical} elongation--response function introduced in Sec.\,\ref{Sec-oscillator-model}. By the formally exact Eqs.\,(\ref{dipole-susc-general})--(\ref{gen-osc-eom}), it is demonstrated that, in case of negligible system memory, the dipole--dipole response function  will acquire the same functional form as the classical response function (Eq.\,(\ref{trig-osc-resp-relax})). If, moreover, a quantum oscillator is chosen as a simple atomic model, the quantum--mechanically determined values for ($\Omega,\Gamma$) turn out to be in perfect agreement with the classical oscillator parameters determined from AL equation.

The intimate relations between quantum--mechanical and classical response and relaxation functions carved out in Sec.\,\ref{Sec-QM} above raise well--founded expectations that the {\sc Lorentz} atom, modelled by Eq.\,(\ref{forced-osc-eq}), will have interesting future applications, in which oscillator parameters are  nowadays determined in quantum--mechanical calculations.
\\*[2em]{\bf\em Acknowledgement:} \\It was my pleasure to discuss  with A. Pelster and J. Akram  many aspects of this work. Special thanks go to V. Bagnato and E. dos Santos  for their warm hospitality and fruitful discussions during a visit to USP Sao Carlos, Brazil, where part of this work developed.

\section{Appendix}
\label{Sec-Appendix}

\subsection{Unique solution of AL equation of motion}
\label{Sec-solution of AL-eom}

The unique solution of Eq.\,(\ref{AL-eq}), which is the general solution of the homogeneous equation plus a particular solution of the inhomogeneous equation, may be cast into the following form ($t\ge t_0$),
\bea
\label{AL-unique-sol-1}
x_{\rm AL}(t,t_0)&=&x_{\rm AL}^{\rm h}(t,t_0)+x_{\rm AL}^{\rm p}(t,t_0)\;,\\
\label{AL-unique-sol-2}
x_{\rm AL}^{\rm h}(t,t_0)&=&\phi(t-t_0)x_0+\ii\chi(t-t_0)m v_0\nonumber\\
&&-\frac{(b_0+\Gamma v_0+\Omega^2 x_0)\tau^2}{1+4\tilde{\Omega}^2 \tau^2}\Big[-\e^{(t-t_0) (\Gamma+1/\tau)}\nonumber\\
&&+\phi(t-t_0)+(\Gamma+1/\tau)\ii\chi(t-t_0)m\Big]\\
\label{AL-unique-sol-3}
x_{\rm AL}^{\rm p}(t,t_0)&=&\int_0^{t-t_0}\de t'~\Big[-\e^{t' (\Gamma+1/\tau)}+\phi(t')\nonumber\\
&&+(\Gamma+1/\tau)\ii\chi(t')m\Big]\frac{\tau f(t-t')}{(1+4\tilde{\Omega}^2 \tau^2)m}\;.
\eea
Here oscillator relaxation and response functions, $\phi(t)$ and $\chi(t)$, and frequency $\tilde{\Omega}$, are defined in terms of $(\Omega,\,\Gamma)$ and are given, resp., in Eq.\,(\ref{osc-resp-relax}) and Eq.\,(\ref{def-om-tilde}). The oscillator parameters $\Omega\g\Omega(\tau,\omega_0)$ and $\Gamma\g\Gamma(\tau,\omega_0)$ are given in terms of the AL parameters $(\tau,\omega_0)$ in Eq.\,(\ref{Gamma-Omega-expanded}).

As discussed in Sec.\,\ref{Sec-no-steady-state} above, Eq.\,(\ref{AL-unique-sol-2}) implies that the unique solution of Eq.\,(\ref{AL-eq}) for initial values $(x_0,\,v_0,\,b_0)$  will diverge, if $(t-t_0)\to\infty$, because the characteristic polynomial of Eq.\,(\ref{AL-eq}) has a positive root, $z_2=\Gamma+1/\tau>0$. From $\lim_{t_0\to-\infty}x_{\rm AL}(t,t_0)=\infty$, I conclude that a steady--state solution of the AL equation does not exist. A steady--state solution would require that $x_{\rm AL}^{\rm h}(t,-\infty)=0$ for generic $(x_0,\,v_0,\,b_0)$.

\subsection{Fourier--Laplace transform (FLT)}
\label{Sec-FLT}

In Eq.\,(\ref{FT-osc-steady-state}), the {\sc Fourier--Laplace} transform (FLT) of a bounded function $f(t)$  ($|f(t)|\le M<\infty$) has been introduced,
\1
\label{def-FLT}
\tilde{f}(z)=\int_{-\infty}^\infty\de t~\e^{\ii t z}\,\ii s\Theta(s t)\, f(t)\;,~~~~s={\rm sign}\Im[z]\ne0\;,
\2
which is an analytical function for all complex $z$ {\em outside} the real axis. The FLT of $f(t)$ has as a {\sc Cauchy}--integral representation
\1
\tilde{f}(z)=\int_{-\infty}^\infty\frac{\de\omega}{\pi}~\frac{f''(\omega)}{\omega-z}
~~\stackrel{z\g\omega\pm\ii o}{\longrightarrow}~~f'(\omega)\pm\ii f''(\omega)
\2
with $f''(\omega)=\frac{1}{2\ii}\left[\tilde{f}(\omega+\ii o)-\tilde{f}(\omega-\ii o)\right]$ denoting the {\em spectral function}, or dissipative part of $\tilde{f}(\omega+\ii o)$, and $f'(\omega)=\frac{1}{2}\left[\tilde{f}(\omega+\ii o)+\tilde{f}(\omega-\ii o)\right]$ denoting the {\em reactive part} of $\tilde{f}(\omega+\ii o)$. Dissipative and reactive parts  obey dispersion relations,
\1
\label{Kramers-Kronig}
f'(\omega)=\pint_{-\infty}^\infty\frac{\de\bar{\omega}}{\pi}~\frac{f''(\bar{\omega})}{\bar{\omega}-\omega}\;,~~
f''(\omega)=-\pint_{-\infty}^\infty\frac{\de\bar{\omega}}{\pi}~\frac{f'(\bar{\omega})}{\bar{\omega}-\omega}\;,
\2
known as {\sc Kramers--Kronig} relations in physics literature.

In general, $f'(\omega)$ and $f''(\omega)$ will be complex functions of the real variable $\omega$. Functions $f(t)$, which  vanish for large $|t|$ (as is the case for response and relaxation functions discussed above), are related to their spectral function by conventional {\sc Fourier} transform,
\1
f''(\omega)=\int_{-\infty}^\infty\frac{\de t}{2} ~\e^{\ii t\omega} f(t)\;,~~f(t)=\int_{-\infty}^\infty\frac{\de\omega}{\pi} ~\e^{-\ii t\omega} f''(\omega)\;,
\2
and one easily verifies for the response function $\chi(t)$ (Eq.\,(\ref{osc-resp-relax})), which is purely imaginary, odd in $t$, and vanishing for $|t|\to\infty$,
\1
\chi''(\omega)=\pm\Im[\tilde{\chi}(\omega\pm\ii o)]=-\chi''(-\omega)=\chi''(\omega)^*\;,
\2
a spectral function which is real, odd in $\omega$, and $1/2$ of the conventional {\sc Fourier} transform of $\chi(t)$.
Similarly, the relaxation function $\phi(t)$ (Eq.\,\ref{osc-resp-relax}), which is real, even in $t$, and vanishing for $|t|\to\infty$, will have a spectral function,
\1
\phi''(\omega)=\pm\Im[\tilde{\phi}(\omega\pm\ii o)]=\phi''(-\omega)=\phi''(\omega)^*
\2
which is real, even in $\omega$, and just $1/2$ of the conventional {\sc Fourier} transform of $\phi(t)$. For response and relaxation spectrum, {\sc Kubo}'s identity takes the simple form: $\chi''(\omega)=\omega\phi''(\omega)/(m\Omega^2)$.

\bibliographystyle{unsrt}
\end{document}